\documentclass[preprint,12pt,showpacs,aps,nofootinbib]{revtex4}
\usepackage{amsfonts}
\usepackage{mathrsfs}
\usepackage{graphicx}
\usepackage{amssymb}
\usepackage{color}
\usepackage{amsmath}
\usepackage{CJK}

\begin{document}
\begin{CJK}{GBK}{kai}

\title{Wilson Line Response of Holographic Superconductors in Gauss-Bonnet Gravity}
\author{Rong-Gen Cai\footnote{Email: cairg@itp.ac.cn}}
\author{Li Li\footnote{Email: liliphy@itp.ac.cn}}
\author{Li-Fang Li\footnote{Email: lilf@itp.ac.cn}}
\author{Hai-Qing Zhang\footnote{Email: hqzhang@itp.ac.cn}}
\author{Yun-Long Zhang\footnote{Email: zhangyl@itp.ac.cn}}
\affiliation{State Key Laboratory of Theoretical Physics,
Institute of Theoretical Physics, Chinese Academy of
Sciences, P.O. Box 2735, Beijing 100190, China}

\begin{abstract}
We study the Wilson line response in the holographic superconducting phase transitions in Gauss-Bonnet gravity. In the black brane background case, the Little-Parks periodicity is independent of the Gauss-Bonnet parameter, while in the AdS soliton case, there is no evidence for the Little-Parks periodicity. We further study the impact of the Gauss-Bonnet term on the holographic phase transitions quantitatively. The results show that such quantum corrections can effectively affect the occurrence of the phase transitions and the response to the Wilson line.
\end{abstract}

\maketitle

\section{Introduction}

The AdS/CFT correspondence~\cite{Maldacena:1997re,Gubser:1998bc,Witten:1998qj} provides us a novel method to study the strongly coupled field
theories through a weakly coupled gravitational system in one higher dimension. In recent years, the AdS/CFT correspondence has been used to mimic many basic phenomena in condensed matter physics, such as Nernst effect~\cite{Hartnoll:2007ih,Hartnoll:2008hs}, superconductivity~\cite{Hartnoll:2008vx,Gubser:2008wv,Nishioka:2009zj} and non-fermi liquid~\cite{Lee:2008xf,Liu:2009dm,Cubrovic:2009ye}. Among these works, the holographic superconductivity is the most studied subject, and  various aspects of the holographic superconductors have been intensively discussed, such as the response to the magnetic field~\cite{Nakano:2008xc,Maeda:2008ir,Montull:2009fe,Cai:2011tm}, the Meissner effect~\cite{Ammon:2009fe,Ammon:2008fc}, the behavior of the entanglement entropy~\cite{Albash:2012pd,Cai:2012sk,Cai:2012nm,Cai:2012es} and the Josephson junction~\cite{Horowitz:2011dz,Wang:2011rva,Kiritsis:2011zq,Wang:2012yj}. For more details, please refer to~\cite{Hartnoll:2009sz,McGreevy:2009xe,Sachdev:2012dq} and references therein.

The Little-Parks (LP) effect was discovered in experiments with empty and thin-walled superconducting cylinders subjected to a parallel magnetic field~\cite{little62}. It states that certain thermodynamic quantities such as the critical temperature $T_c$ are periodic functions of the enclosed magnetic flux with period $h c/2e$ where $e$ is the fundamental charge of carriers. While recently it was shown in the condensed matter literatures~\cite{Tesanovic,vakaryuk08,loder08,wei08,barash08} that the LP period would be broken for small enough cylinder, the LP degeneracy and its uplifting have been realized holographically in~\cite{montull11}, where the Wilson line $W$ along the compact direction is identified as a quantum hair. They found that the boundary field theory in a deconfined state dual to black brane is insensitive to the quantum hair, thus the Aharonov-Bohm (AB) effect is suppressed and the LP period appears. In contrast, the theory in the confining vacuum dual to the soliton is sensitive to the quantum hair, thus the AB effect is unsuppressed and there is no trace of the LP effect.

This encouraging discovery motivates us to study the responses to a Wilson line on a circle in the frame of Gauss-Bonnet gravity.
In~\cite{montull11,montull12}, the effect of the Wilson line has been studied in the Einstein gravity
background. It would be interesting to see how the modification of the bulk gravity may influence the response
to the Wilson line. The higher curvature corrections originated from the string theory can effectively describe quantum corrections in the bulk. According to the AdS/CFT correspondence, such effects on the gravity side map to $1/\mathcal{N}$ suppressed corrections in the large $\mathcal{N}$ expansion of the boundary field theory. The studies of holographic superconductor with higher curvature
corrections~\cite{Gregory:2009fj,Pan:2009xa,Cai:2010zm,Kuang:2010jc,Siani:2010uw,Wu:2010vr} indicate that such terms can quantitatively affect many properties of the boundary system. In particular, the ratio of the frequency gap over the critical temperature, i.e., $\omega_g/T_c$ which was found to be universal in Einstein gravity, breaks down. Therefore, in this paper, we are interested in how the Gauss-Bonnet term changes the response to the Wilson line. We study the responses in both AdS black brane background which can mimic the holographic conductor/superconductor transition, and AdS soliton background which can model the holographic insulator/superconductor transition. We work in the probe limit. Our results show that the responses to Wilson line are dramatically different between holographic conductor/superconductor phase transition and holographic insulator/superconductor phase transition. Explicitly, for black brane, the LP periodicity still holds in the Gauss-Bonnet gravity and is independent of the Gauss-Bonnet parameter, while there is no evidence for the LP periodicity in the AdS soliton case. These phenomena are similar to those in the Einstein theory. Furthermore, we analyze the impact of the Gauss-Bonnet term to the systems in detail. We find that with the change of the Gauss-Bonnet parameter, the corresponding physical quantities, such as the condensation and the current, display regular behaviors. In particular, at fixed chemical potential, the behaviors of condensation and current with respect to $a_\chi$ ( or equivalently Wilson line $W$) are much more different in the two systems with the change of the Gauss-Bonnet parameter.

Our paper is organized as follows. First, we construct the holographic model in Section.~\ref{Sec.2}. We study the responses to the Wilson line in the holographic conductor/superconductor model and the insulator/superconductor model in Section.~\ref{Sec.3} and Section.~\ref{Sec.4}, respectively.
In the last Section.~\ref{Sec.5}, we give some conclusions and discussions.

\section{The holographic model}\label{Sec.2}
Let us start from the Einstein-Gauss-Bonnet gravity  with a negative cosmological constant coupled to a $U(1)$ gauge field $A_{\mu}$ and a charged scalar field $\Psi$ in 5-dimensional spacetime. The action reads
\begin{equation}\label{action}
S=\int d^5 x \sqrt{-g} \Big[\frac{1}{2
\kappa_5^2}\Big(R+\frac{12}{L^2}+\frac{\alpha}{{2}} (R^2-4R^{\mu\nu}
R_{\mu\nu}+R^{\mu\nu\rho\sigma}R_{\mu\nu\rho\sigma})\Big){+}\frac{1}{\hat{g}^2}\Big(-\frac{1}{4}F^2_{\mu\nu}-\frac{1}{L^2}|D_{\mu}\Psi|^2\Big)\Big],
\end{equation}
with $\kappa_5$ the gravitational constant, $L$ the radius of the AdS spacetime, $F_{\mu\nu}=\partial_\mu A_\nu-\partial_\nu
A_\mu$ and $D_{\mu}=\partial_{\mu}-iA_{\mu}$. The quadratic curvature term is the Gauss-Bonnet term with $\alpha$ the Gauss-Bonnet parameter.
In order to compare with the results in~\cite{montull11,montull12}, we only consider the case with a massless scalar field. To mimic a boundary system compactified on a circle, we are interested in the geometry with one compact spatial direction labeled as $\chi$ with $0\leq\chi<2\pi R$. In this paper, we will work at finite temperature $T$, corresponding to a compact Euclidean time direction with radius $\beta=1/T$.

The control parameter that we will consider in the holographic superconductor is a Wilson line
along the compact direction $\chi$, with a constant non-trivial gauge vector potential $a_\chi$.
\begin{equation}
W=\exp\left(ei\oint dx^\mu a_\mu\right)=\exp\left(i e 2\pi R a_\chi\right),
\end{equation}
where the integral is calculated along the compact direction and $e$ is the fundamental charge. The Wilson line on the material can be thought to be generated by the axial magnetic flux since the circulation of the gauge potential equals the magnetic flux enclosed by the path.
One parameter that can characterize the response of $W$ is the fluxoid number
\begin{equation}
m\equiv\frac{1}{2\pi}\oint dx^{\mu}\partial_{\mu}\theta=\mathrm{integer},
\end{equation}
where the integral is done along the compact direction and $\theta$ is the phase of the order parameter.

In general, we should solve the full coupled equations of motion, which are more complicated in the case with the Gauss-Bonnet term. However, we can get some qualitative features in the so-called probe limit. Indeed, we can see from the action that in the limit $\kappa_5^2/\hat{g}^2\ll 1$, the back reaction of the gauge field and the complex scalar field can be neglected safely. For our case, the simplest ansatz are as follows
\begin{equation}\label{field_equation}
\psi=\psi(z)e^{i m\chi/R}, \ \ \ A_t=A_t(z), \ \ \ A_{\chi}=A_{\chi}(z).
\end{equation}
Near the AdS boundary $z\rightarrow0$, the scalar field and the Maxwell field behave as
\begin{eqnarray}
\psi&=&s+ \langle O \rangle z^4/4+\cdots,\nonumber\\
A_t&=&\mu-\rho z^2/2+\cdots,\nonumber\\
A_{\chi}&=&a_{\chi}+J_{\chi}z^2/2+\cdots.\label{boundary}
\end{eqnarray}
From the AdS/CFT dictionary, the coefficients above can be related to physical quantities in the boundary field theory. $\langle O \rangle$ is the vacuum expectation value (VEV) of the dual operator with $s$ the source which is set to be zero to accomplish the spontaneous symmetry breaking of the gauge symmetry. $\mu$ and $\rho$ are chemical potential and charge density, respectively. $J_{\chi}$ is the VEV of the $U(1)$ current and $a_{\chi}$ plays the role of a gauge potential along the compact direction of the boundary material. The free energy $F=T S_{E}$ can be obtained from the AdS Euclidean action $S_E$ evaluated with all the bulk fields on shell, which is used to determine which configuration is thermodynamically favorable.

In the next two sections, we will study the response to the Wilson line in both five dimensional Gauss-Bonnet-AdS black brane and Gauss-Bonnet-AdS soliton backgrounds. Especially, we concentrate on the quantitative changes of boundary systems after turning on such quadratic curvature corrections in the bulk.

\section{Holographic conductor/Superconductor transition}\label{Sec.3}
The 5-dimensional Gauss-Bonnet-AdS black brane with a Ricci flat
horizon is described by~\cite{Cai}
\begin{eqnarray}
ds^2_{BB}
&=&\frac{L^2}{z^2}\Big[-f(z)dt^2+dx^2+dy^2+d\chi^2+\frac{dz^2}{f(z)}\Big],
\\
f(z) &=&\frac{L^2}{2\alpha}\bigg(1-\sqrt{1-\frac{4\alpha}{L^2}\big(
1-\frac{z^4}{z_h^4}\big)}\bigg), \nonumber
\end{eqnarray}
where the horizon is located at $z_h$. The temperature of the black brane is $T=\frac{1}{\pi z_h}$. In order to have a well-defined vacuum for the gravity theory, one has to have $\alpha\leq L^2/4$. The upper bound $\alpha =L^2/4$ is called the Chern-Simons limit. If we further consider the causality constraint from the boundary CFT, there is an additional constraint on the Gauss-Bonnet parameter with $-7L^2/36\leq \alpha \leq 9L^2/100$~\cite{brigante07,Brigante:2008gz,Buchel:2009tt,Hofman:2009ug}. In the AdS/CFT correspondence, the temperature of the black hole is just the one of the dual field theory. In the following
numerical calculations we will set $L=1$.

Assuming the matter field in the form of Eq.(\ref{field_equation}), we can obtain the equations of motion
\begin{eqnarray}\label{BB_motion}
z^{3}\partial_{z}\Big(\frac{f}{z^{3}}\partial_{z}\psi\Big)+\Big[\frac{A_t^2}{f}-\Big(A_{\chi}-m/R\Big)^2\Big]\psi=0,\nonumber\\
z\,\partial_z\Big(\frac{\partial_z A_t}{z}\Big)-\frac{2A_t}{z^2f}\psi^2=0,\nonumber\\
z\,\partial_z\Big(\frac{f\partial_z A_{\chi}}{z}\Big)-\frac{2\Big(A_{\chi}-m/R\Big)}{z^2}\psi^2=0.
\end{eqnarray}
To solve the above equations of motion, we impose the regularity conditions at the horizon $z=z_h$
\begin{eqnarray}\label{BB_boundary}
\frac{4}{z_h}\partial_z\psi+(A_{\chi}-m/R)^2\psi=0,\nonumber\\
A_t=0,\nonumber\\
\partial_z A_{\chi}+\frac{1}{2z_h}(A_{\chi}-m/R)\psi^2=0.
\end{eqnarray}
From the bulk equations of motion Eq.(\ref{BB_motion}) and the boundary conditions Eq.(\ref{BB_boundary}), we see that $A_{\chi}$ equivalently appears in the combination $(m/R-A_{\chi})$ which comes from the local covariant quantity $D_{\chi}\psi=i(m/R-A_{\chi})\psi e^{im\chi/R}$. This implies that the effective action of the boundary system will only depend on local gauge invariant quantities, and thus will display LP periodicity. This observation is confirmed by our subsequent numerical calculations.

Considering the scaling symmetry
\begin{equation}
(z,t,x,y,\chi)\rightarrow\lambda(z,t,x,y,\chi),\ \  R\rightarrow\lambda R,\ \  A_{\chi}\rightarrow A_{\chi}/\lambda,\ \ A_t\rightarrow A_t/\lambda,
\end{equation}
we can adjust the solutions to satisfy $z_h=1$. And the corresponding scaling invariant variables are $a_{\chi}R$, $\langle O\rangle R^4$, $J_{\chi}R^3$ and so on. The free energy of the system is
\begin{eqnarray}\label{free_erengy_a}
F/V_{3}=-\frac{1}{2}(\rho\mu+a_{\chi}J_{\chi})-\int_{0}^{z_h}dz\frac{\psi^2}{z^3}\Big[-\frac{A_t^2}{f(z)}+A_{\chi}\big(A_{\chi}-\frac{m}{R}\big)\Big],
\end{eqnarray}
which is used to determine which configuration is thermodynamically favored, where $V_3$ is the spatial volume of the black hole spanned by $x$, $y$ and $\chi$.
\begin{figure*}[h!]
\begin{center}
\includegraphics[width=0.45\textwidth]{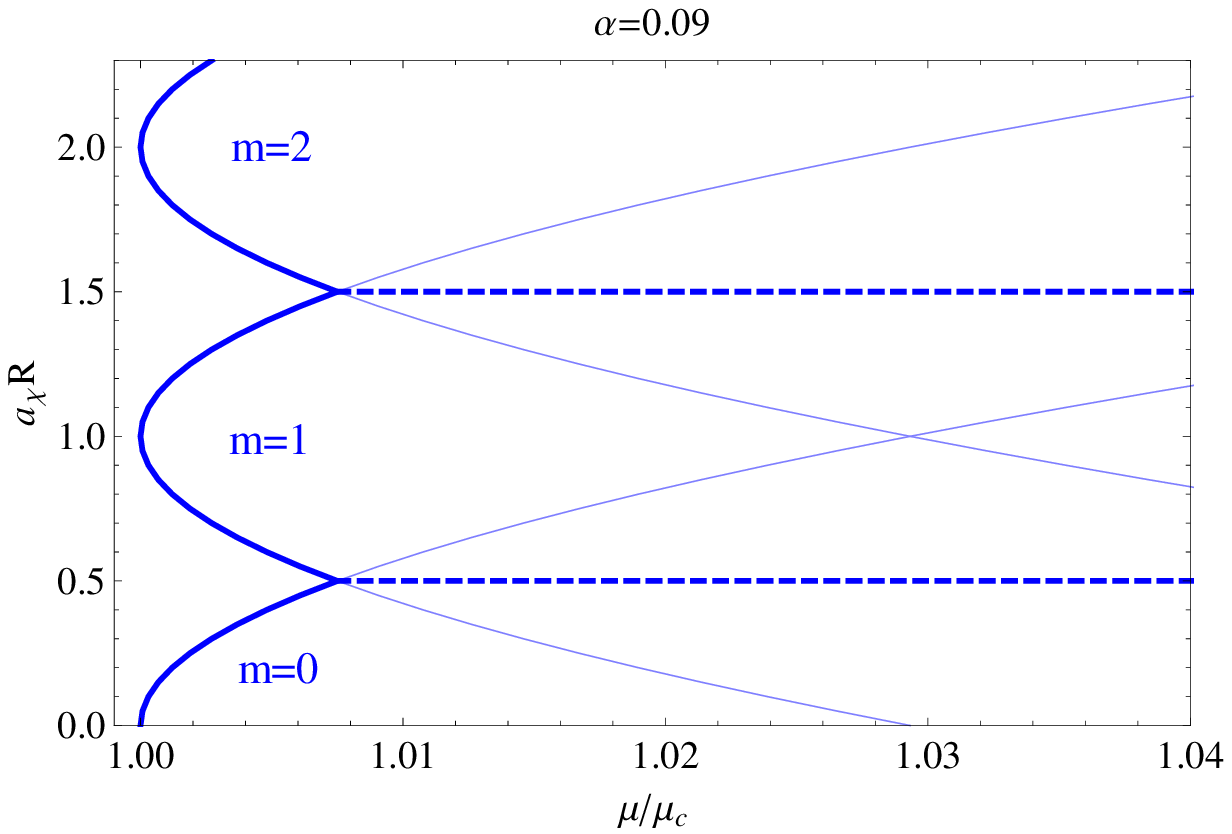}
\includegraphics[width=0.45\textwidth]{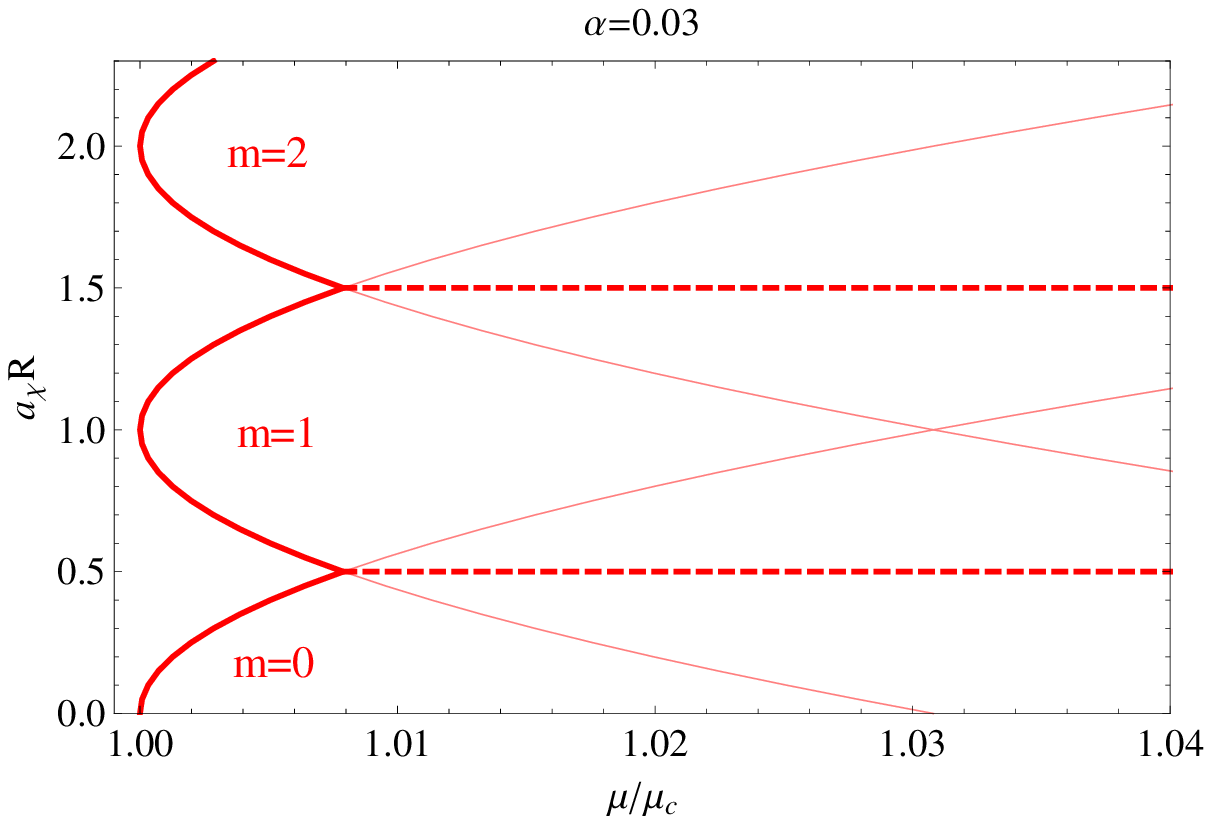}\\
~~\\
\includegraphics[width=0.45\textwidth]{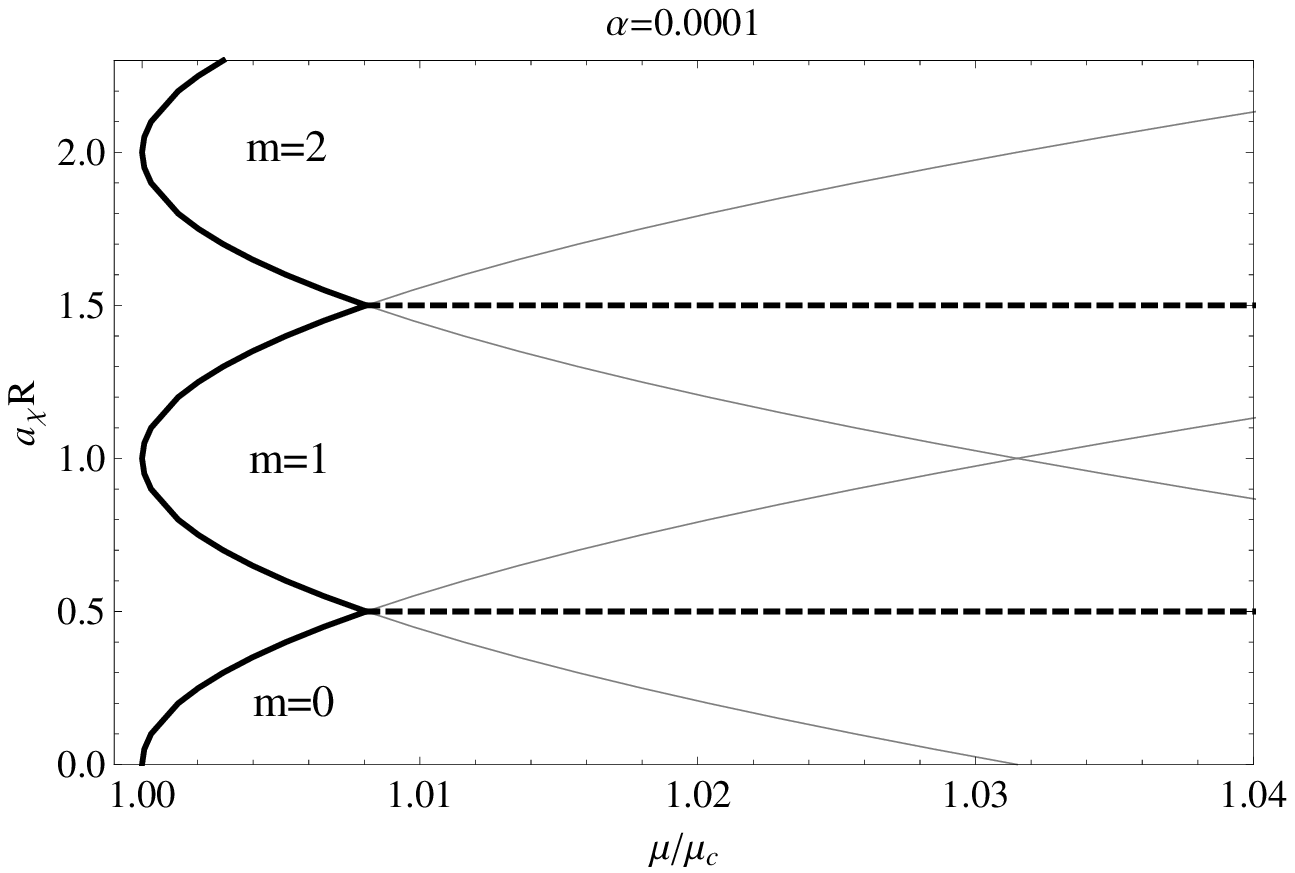}
\includegraphics[width=0.45\textwidth]{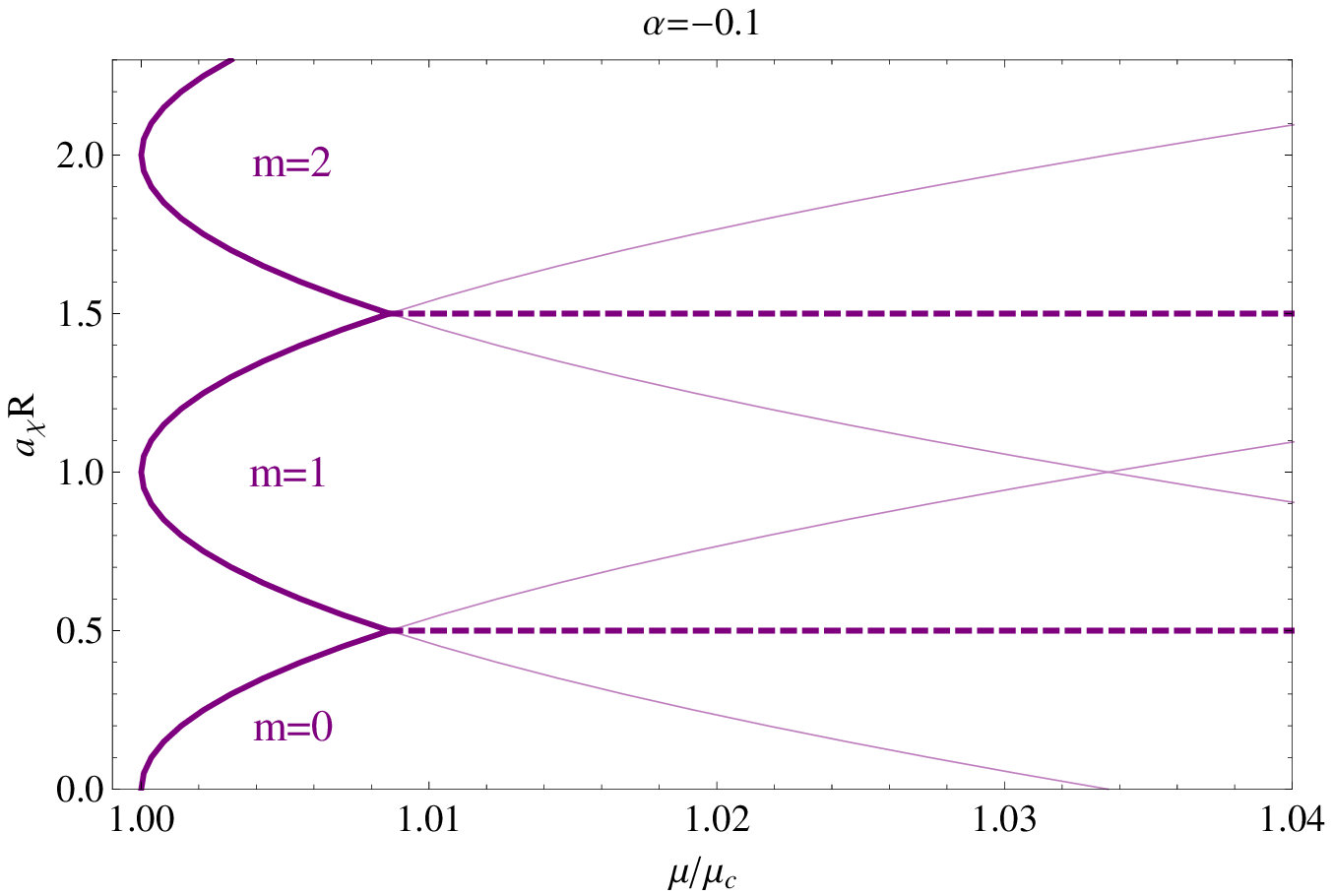}
\caption{{(Color Online)} Phase diagrams for the conductor/superconductor transition at $T=1/\pi R$ for different Gauss-Bonnet parameter $\alpha$. Thick solid lines represent the phase boundary between superconducting and normal phase. Thin solid lines label the existence of the m-fluxoid condensates. Dashed lines separate different fluxoid domains. Here $\mu_c$ is the critical chemical potential when $a_{\chi} R=0$ (without Wilson line), which is different for different $\alpha$.}\label{fig1}
\end{center}
\end{figure*}

\begin{figure}[h!]
\begin{center}
\includegraphics[width=0.7\textwidth]{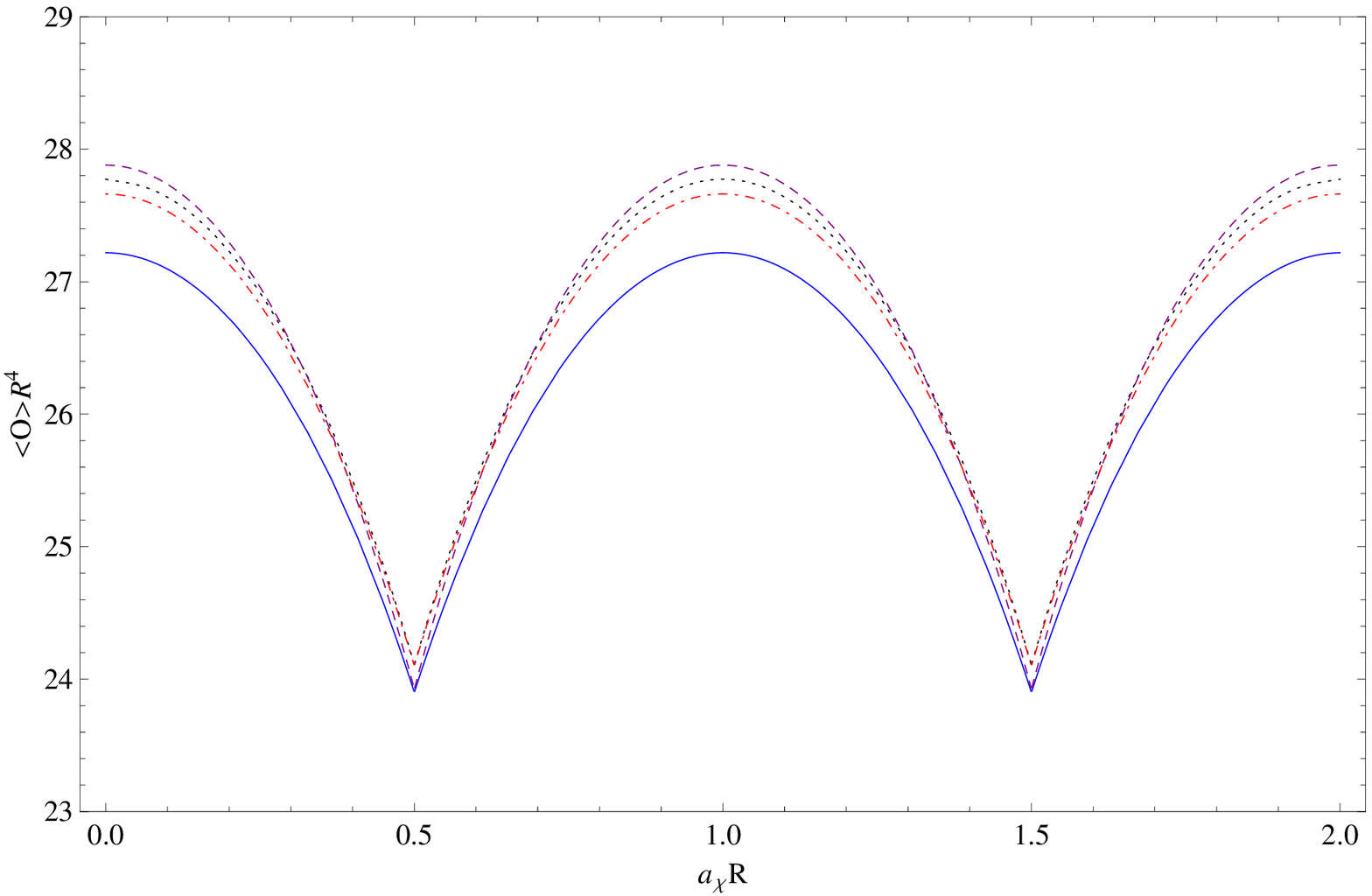}
\caption{(Color Online). The modulus of the condensate $\langle
O\rangle$ as a function of $a_{\chi}$ at $\mu=1.03\mu_c$. Different
color curves with period represent the value of condensation for
different $\alpha$. From the top to the bottom, the curves with
different colors correspond to $\alpha=-0.1$, $\alpha=0.0001$,
$\alpha=0.03$, and $\alpha=0.09$, respectively.}\label{fig3}
\end{center}
\end{figure}
Fig.\ref{fig1} shows the phase diagrams for the occurrence of superconductivity in the Gauss-Bonnet-AdS black brane case. Here we choose the Gauss-Bonnet parameter as $\alpha=0.09$, $\alpha=0.03$, $\alpha=0.0001$, and $\alpha=-0.1$ in turn. Although we introduce some kind of quantum corrections effectively described by Gauss-Bonnet term in the bulk, we can see from Fig.\ref{fig1} that the phase diagram displays a precise periodicity with period $\Delta a_{\chi}R=1$ no matter the choice of the Gauss-Bonnet parameter $\alpha$. When the parameter $\alpha$ vanishes, the system returns to the one in Einstein theory. Following the terminology in~\cite{montull11}, we conclude that the AB effects are suppressed for black brane background. However, the introduction of Gauss-Bonnet term can quantitatively affect other physical observables. For different $\alpha$, $\mu_c$ is different accordingly, where $\mu_c$ is the critical chemical potential denoting the occurrence of phase transition when $a_{\chi}=0$ (without Wilson line). For $\alpha = 0.09$, $\alpha = 0.03$, $\alpha = 0.0001 $, and $\alpha = -0.1$, the corresponding critical chemical potentials are $\mu_c R \approx 7.13$, $\mu_c R \approx 6.69$, $\mu_c R \approx 6.50$, and $\mu_c R \approx 6.02$, respectively. It is clear that as one increases the Gauss-Bonnet parameter $\alpha$, $\mu_c$ increases accordingly, which makes the condensation more difficult to form for vanishing magnetic flux.

\begin{table}[h]
\caption{\label{table1} The coefficients $k_{\alpha}$ for different
Gauss-Bonnet parameters $\alpha$.}
\begin{center}
\begin{tabular}{cccccc}
 \hline
 $~~  ~~$ & $~~\alpha=0.09~~$ & $~~\alpha=0.03~~$ & $~~\alpha=0.0001~~$  & $~~\alpha=0~~$ & $~~\alpha=-0.1~~$ \\
 \hline
 $k_{(\alpha)}$ & $0.1727$ & $0.1771$ & $0.1790$ & $0.1791$ & $0.1850$ \\
 $k_{(\alpha)}/k_{(\alpha=0)}$ & $0.9642$ & $0.9888$ & $0.9999$ & $1.0000$ & $1.0334$ \\
 \hline
\end{tabular}
\end{center}
\end{table}
%
\begin{figure}[h!]
\begin{center}
\includegraphics[width=0.7\textwidth,viewport=0 100 500 400,clip=true]{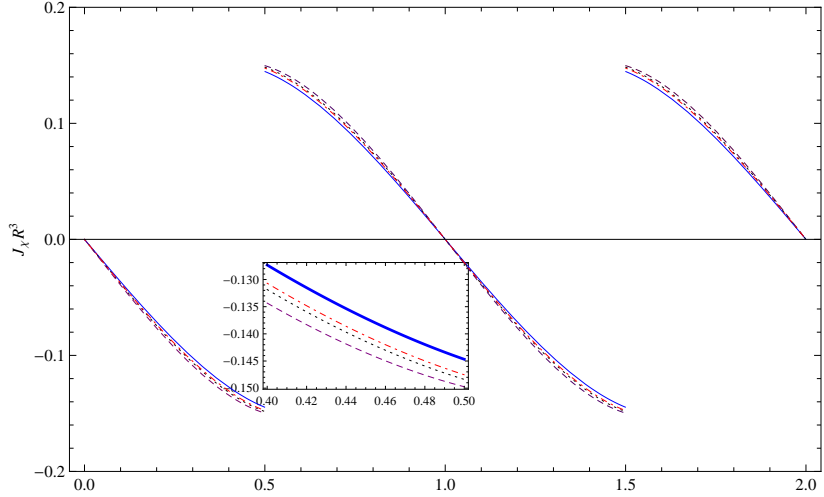}
\caption{(Color Online). The modulus of the current $J_{\chi}$ as a
function of $a_{\chi}$ at $\mu=1.03\mu_c$. For the first half
period, from the top to the bottom, the values of $\alpha$ are
$\alpha=0.09$ (blue solid line), $\alpha=0.03$ (red dot-dashed
line), $\alpha=0.0001$ (black dotted line), $\alpha=-0.1$ (purple
dashed line), respectively. Details can be seen in the
context.}\label{fig4}
\end{center}
\end{figure}
Comparing with the above four phase diagrams, we can see that, for a given $a_{\chi}R$, the smaller the Gauss-Bonnet parameter $\alpha$ is chosen, the bigger $\mu/\mu_c$ is required to trigger the phase transition. The variation of the critical chemical potential to the gauge potential and fluxoid number $m$ is found to perfectly behave as
\begin{equation}
1-({\mu}/{\mu_c})^{-1}=k_{(\alpha)}^2(a_{\chi}R-m)^2,
\end{equation}
where $\mu$ is the critical chemical potential for non-vanishing $a_{\chi}R$. The coefficients $k_{(\alpha)}$ corresponding to different $\alpha$ are listed in Table.~\ref{table1}.

According to~\cite{little62,wei08}, the coefficient $k_{(\alpha)}$ would reflect the ratio of the coherence length to the cylinder radius,
i.e. $k_{(\alpha)}\sim \xi_0/R$.~\footnote{Strictly speaking, there is a little subtle to interpret $k_{(\alpha)}\sim\xi_0/R$. The result in \cite{wei08} is obtained in terms of the Gor'kov approach to BCS theory, while our system is in strongly coupled region. However, this observed behavior here is quite similar to that one and  therefore we adopt the terminology to describe this behavior.} Thus, from Table.~\ref{table1}, we may conclude that compared to the Einstein case, the positive Gauss-Bonnet corrections would decrease the coherence length $\xi_0$ of the Cooper-pair, while the negative corrections increase the coherence length $\xi_0$. Since $\xi_0$ plays the role of the inverse of the mass of the pair, we can find that the quantum corrections described by Gauss-Bonnet term can change the effective mass of the charge carriers.

Fig.\ref{fig3} shows  the condensation via changing $a_\chi\sim
\ln(W)$ for fixed chemical potential $\mu=1.03\mu_c$. From the top
to the bottom, $\alpha$ increases. This implies that with the
decrease of the Gauss-Bonnet parameter $\alpha$, the condensation
$\langle O\rangle$ becomes bigger. Fig.~\ref{fig4} shows us the
change of the current $J_{\chi}$ as a function of $a_{\chi}$ at
fixed chemical potential $\mu=1.03\mu_c$. The behavior of the
current is slightly different for different $\alpha$. When $\alpha$
decreases, the line becomes more inclined. That is to say, for a
given $a_{\chi}$, the magnitude of the current increases with the
decrease of $\alpha$. This phenomenon reflects that smaller $\alpha$
is more sensitive to the response to Wilson line.

Therefore, from the above diagrams, we clearly see that the LP period $\Delta a_{\chi}=1/R$ still holds in Gauss-Bonnet gravity no matter the choice of  the Gauss-Bonnet parameter, and also the period is the same as that in Einstein theory~\cite{montull12}. In the same spirit of~\cite{montull12}, the existence of LP period would be understood by that the effective action of the boundary system has no direct dependence on non-local gauge invariants such as the Wilson line $W$ and fluxoid number $m$. As argued in \cite{montull12}, the existence of LP period implies that the AB effects are also somehow suppressed in our case in the limit $\mathcal{N}\rightarrow \infty$ with $\mathcal{N}$ the number of colors. Although including the Gauss-Bonnet term can not break LP period, this kind of quantum correction does impose its effect on many physical quantities of the dual boundary system. We focus our attention on the impact of different Gauss-Bonnet parameters. Our results show that the Gauss-Bonnet parameter can quantitatively affect the occurrence of phase transition and the response to the Wilson line.

\section{Holographic insulator/Superconductor Transition}\label{Sec.4}
The AdS soliton metric in the Gauss-Bonnet gravity  reads~\cite{Cai:2007wz}
\begin{eqnarray}\label{soliton}
ds^2_{SL}
&=&\frac{L^2}{z^2}\Big[-dt^2+dx^2+dy^2+f(z)d\chi^2+ 
\frac{1}{f(z)}dz^2\Big],
\\
f(z) &=&\frac{L^2}{2\alpha}\bigg(1-\sqrt{1-\frac{4\alpha}{L^2}(1-\frac{z^4}{z_0^4})}\bigg). \nonumber
\end{eqnarray}
Obviously there does not exist any horizon in this soliton solution, but a conical singularity at the tip $z=z_0$, which obeys $f(z_0)=0$. To remove this singularity, the coordinate $\chi$ must have a period $\pi z_0$. This asymptotical AdS solution is dual to a boundary theory in the confining vacuum, which is reminiscent of the insulator phase in condensed matter physics.

The equations of motion for this system are
\begin{eqnarray}
z^{3}\partial_{z}\Big(\frac{f}{z^{3}}\partial_z\psi\Big)+\Big[A_t^2-\frac{(A_{\chi}-m/R)^2}{f}\Big]\psi=0,\nonumber\\
z\,\partial_z\Big(\frac{f\partial_z A_t}{z}\Big)-\frac{2A_t}{z^2}\psi^2=0,\nonumber\\
z\,\partial_z\Big(\frac{\partial_z A_{\chi}}{z}\Big)-\frac{2\Big(A_{\chi}-m/R\Big)}{z^2f}\psi=0.
\end{eqnarray}
At the tip $z=z_0$, the requirement of regularity on the above set of equations implies the following boundary conditions
\begin{eqnarray}
-\frac{4}{z_0}\partial_z\psi+A_t^2\psi=0\ \ \ \mathrm{for} \ \ m&=&0,\ \ \ \psi=0 \ \ \mathrm{for}\ \ m\neq 0,\nonumber\\
\partial_z A_t+\frac{A_t}{2z_0}\psi^2&=&0,\nonumber\\
A_{\chi}&=&0.
\end{eqnarray}

Different from the Gauss-Bonnet black brane case, in the soliton case, the boundary conditions now depend directly on $A_{\chi}$ at $z_0$. Since we demand a Wilson line in the external gauge field $a_\chi\sim \ln(W)$, $A_\chi(z)$ acquires a nontrivial profile which generates a magnetic field $F_{z\chi}$ in the bulk. Furthermore, the requirement $A_{\chi}(z_0)=0$ breaks down the gauge equivalence among different fluxoid sectors. In addition, we can see that the boundary conditions are also sensitive to $m$. These differences between the black brane and soliton geometry may lead to several distortions of the dual system. Similar to black brane case, we can solve the equations of motion numerically and obtain our simulation results. The free energy of such a system is
\begin{eqnarray}\label{free_erengy_a}
F/V_{3}=-\frac{1}{2}(\rho\mu+a_{\chi}J_{\chi})-\int_{0}^{z_0}dz\frac{\psi^2}{z^3}\Big[-A_t^2+\frac{A_{\chi}}{f(z)}(A_{\chi}-\frac{m}{R})\Big],\label{free_energy_b}
\end{eqnarray}
which is used to find the thermodynamically favored fluxoid sector.

\begin{figure}[h!]
\begin{center}
\includegraphics[width=0.45\textwidth]{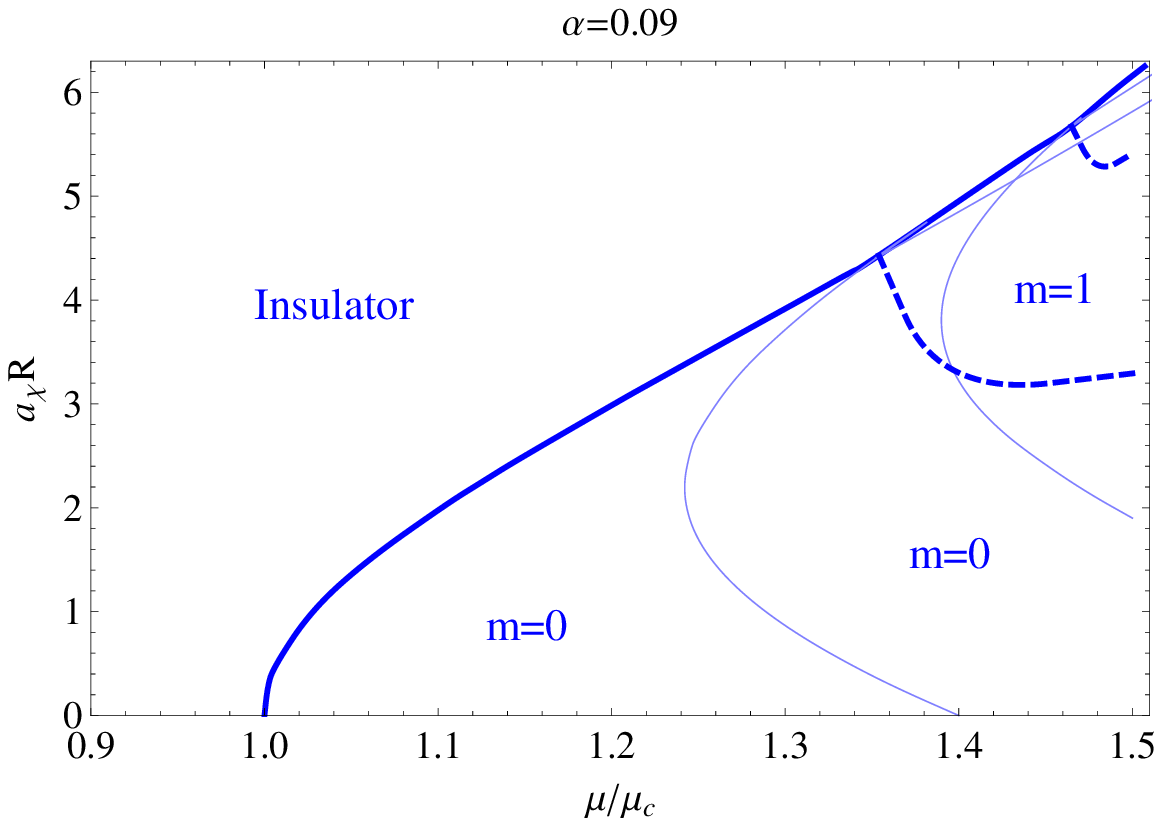}
\includegraphics[width=0.45\textwidth]{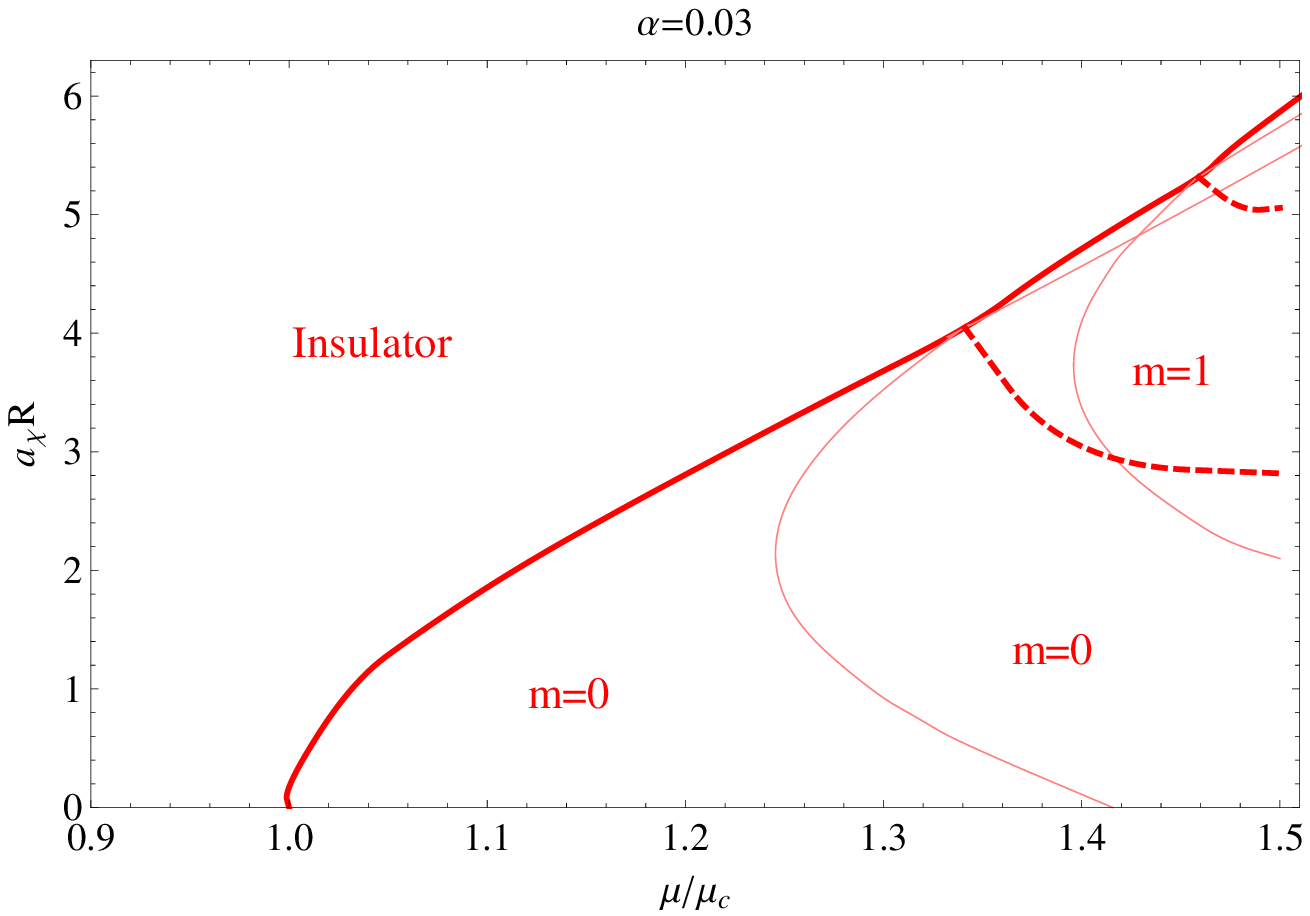}\\
~~\\
\includegraphics[width=0.45\textwidth]{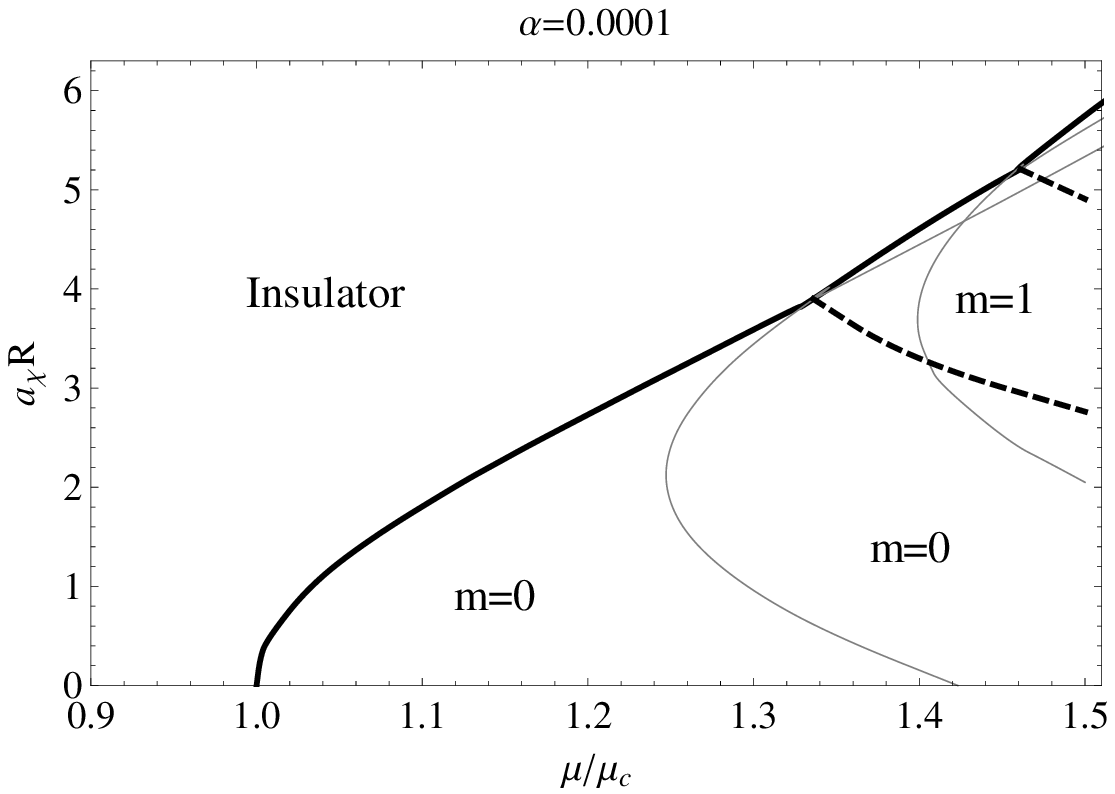}
\includegraphics[width=0.45\textwidth]{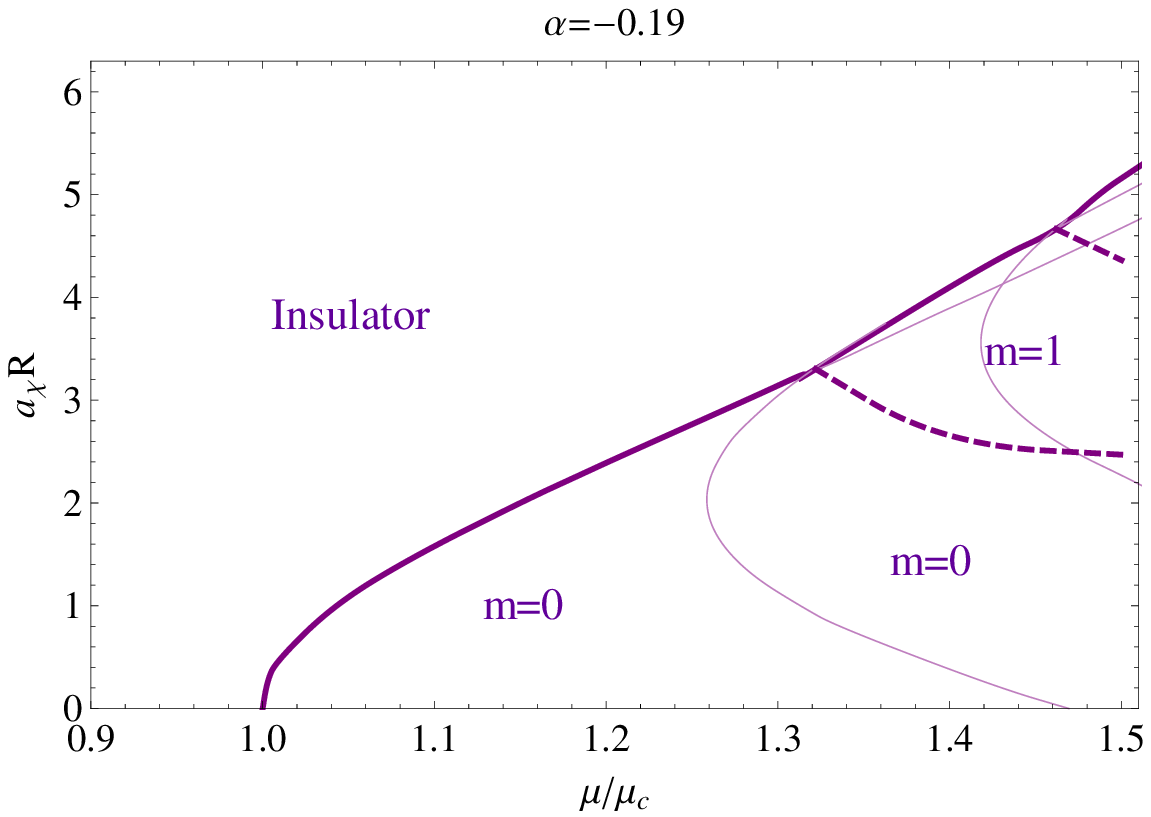}
\caption{(Color Online). Phase diagrams for the soliton background. Each subgraph shows the phase diagram for different Gauss-Bonnet parameter $\alpha$. Thick solid lines represent the phase transition between the superconducting and insulating  phases. Thin solid lines denote the existence of m-fluxoid condensates. Dashed lines separate different fluxoid domains. Here $\mu_c$ is the critical chemical potential with $a_{\chi}=0$ and it  varies for different $\alpha$.}\label{fig5}
\end{center}
\end{figure}

\begin{figure}[h!]
\begin{center}
\includegraphics[width=0.7\textwidth]{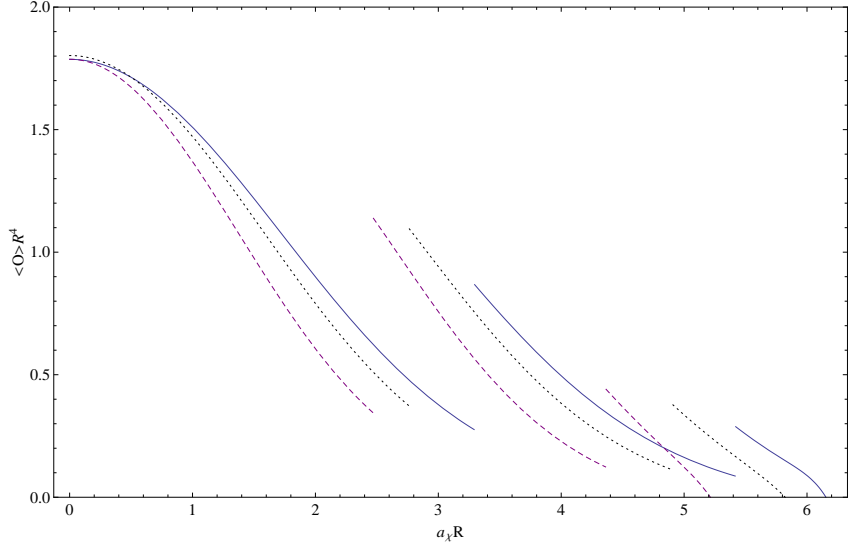}
\caption{(Color Online). The modulus of the condensate $\langle O \rangle$ as a function of $a_{\chi}$ for fixed chemical potential $\mu=1.5\mu_c$. The first jump exists when the $m=1$ solution becomes energetically favorable for every $\alpha$. The second jump occurs when the $m=2$ solution becomes the grand state for every $\alpha$. From the top to the bottom, $\alpha=0.09$ (blue solid line), $\alpha=0.0001$ (black dotted line) and $\alpha=-0.19$ (purple dashed line) in turn. With the increase of $\alpha$, the occurrence of the jump moves to the right.}\label{fig7}
\end{center}
\end{figure}
\begin{figure}[h!]
\begin{center}
\includegraphics[width=0.7\textwidth]{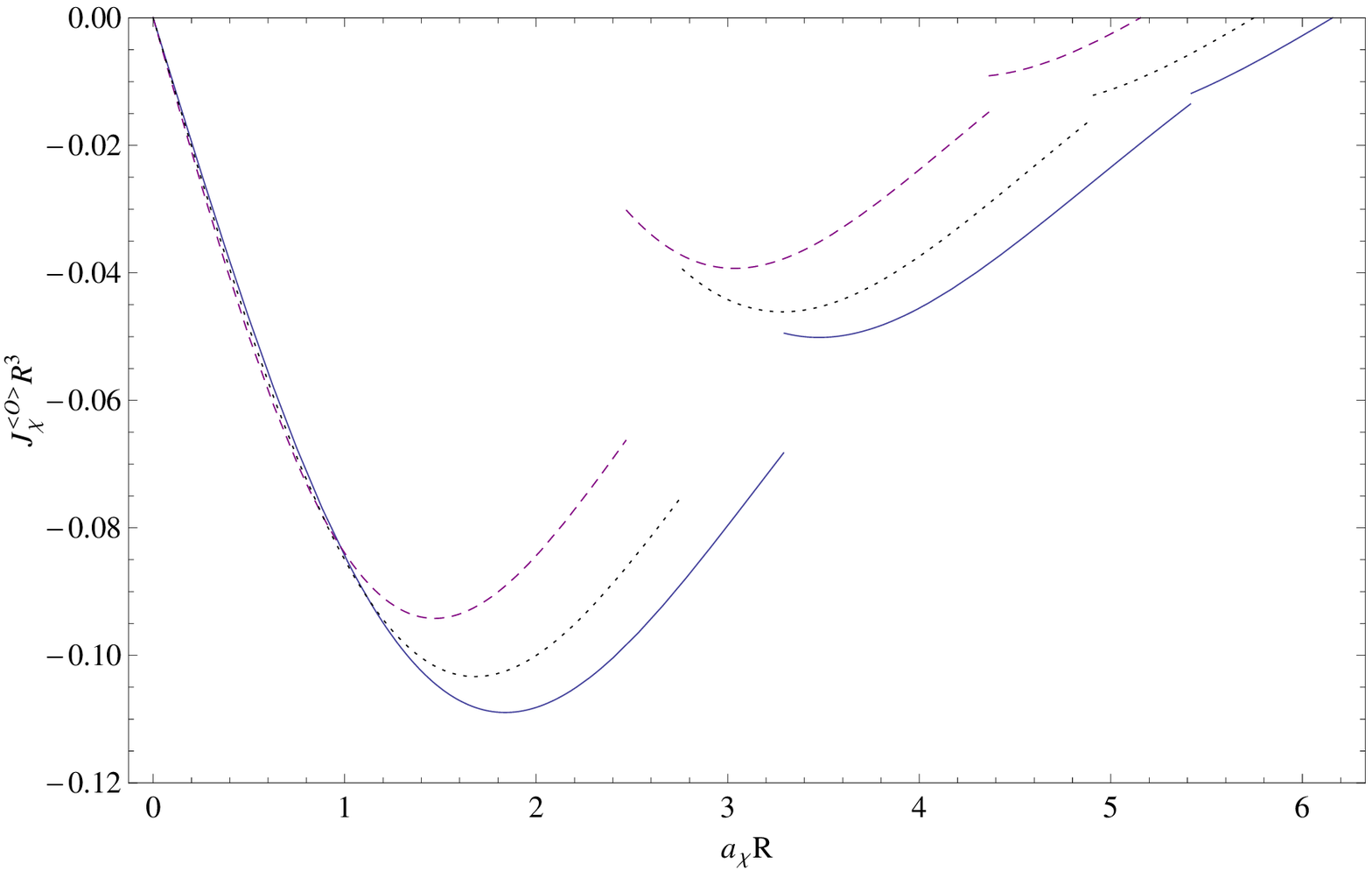}
\caption{(Color Online). The modulus of the current $J_\chi^{\langle O\rangle}$ as a function of $a_{\chi}$ for a fixed chemical potential $\mu=1.5\mu_c$. The first jump exists when the $m=1$ solution becomes energetically favorable for every $\alpha$. The second jump occurs when the $m=2$ solution becomes the grand state for every $\alpha$. From the top to the bottom, $\alpha=-0.19$ (purple dashed line), $\alpha=0.0001$ (black dotted line) and $\alpha=0.09$ (blue solid line) in turn. With the increase of $\alpha$, the occurrence of the jump moves to the right.}\label{fig8}
\end{center}
\end{figure}

Fig.~\ref{fig5} shows us the phase diagrams for the soliton case. Each subgraph corresponds to different $\alpha$. Our results show that the LP period is destroyed in the Gauss-Bonnet-AdS soliton case. In the rest of this section, we study how physical observables are affected as one tunes the strength of the quantum corrections measured by the Gauss-Bonnet parameter $\alpha$.
Here $\mu_c$ is the critical chemical potential when $a_{\chi}$ vanishes. For $\alpha=0.09$, $\alpha=0.0001$ and $\alpha=-0.19$, the corresponding  critical chemical potential is $\mu_c R\approx 1.77$, $\mu_c R \approx 1.70$ and $\mu_c R \approx 1.60$, respectively. We can see that $\mu_c$ increases with the increase of the Gauss-Bonnet parameter, which makes the phase transition from insulator to superconductor more difficult to happen for $a_{\chi}=0$. Compared with each phase diagram for different $\alpha$, it is easy to see that the boundary of the phase diagram gradually drops down as one decreases the Gauss-Bonnet parameter $\alpha$. This implies that at given $\mu/\mu_c>1$, the superconducting phase can be more easily destroyed by the applied $a_\chi$ for smaller $\alpha$.

Fig.~\ref{fig7} shows the evolution of the condensation $\langle O \rangle$ as a function of $a_{\chi}$ at $\mu=1.5\mu_c$ for different $\alpha$. The first jump appears when the $m=1$ sector becomes thermodynamically favorable. The second jump occurs when the $m=2$ sector becomes the ground state for every $\alpha$. The jumps here indicate that the effective Lagrangian of the boundary theory gets the non-trivial dependence on fluxoid number $m$. The position for the jumps between different fluxoid domains
of different $\alpha$ is listed in Table.~\ref{table2}.
\begin{table}[h]
\caption{\label{table2} The position of the jumps for different Gauss-Bonnet parameters}
\begin{center}
\begin{tabular}{cccccc}
 \hline
 $~~  ~~$ & $~~\alpha=0.09~~$ & $~~\alpha=0.0001~~$ & $~~\alpha=-0.19~~$ \\
 \hline
 $a_{\chi}R\mathrm{\ (1st \ jump)}$ & $3.29$ & $2.76$ & $2.46$ \\
 $a_{\chi}R\mathrm{\ (2nd \ jump)}$ & $5.41$ & $4.92$ & $4.36$ \\
 \hline
\end{tabular}
\end{center}
\end{table}
From this table, we can find that with the increase of $\alpha$, the occurrence of the jumps move to the right. Further, if we fix $a_{\chi}R$ in Fig.~\ref{fig7}, we see that the condensation $\langle O \rangle$ is larger for larger $\alpha$. This property implies that with the increase of $\alpha$, the condensation gap becomes higher. This behavior is opposite to the black brane case discussed in the last section.

The behavior of the current $J_{\chi}^{\langle O\rangle}\equiv J_{\chi}-J_{\chi}^{vac}$ as a function of $a_{\chi}$ at fixed $\mu/\mu_c$ is shown in Fig.~\ref{fig8}. $J_{\chi}$ is the total current in superconducting phase. $J_{\chi}^{vac}$ is a normal-phase persistent-current presenting in the soliton (superconductor or not), which can be read from normal phase for vanishing $\psi$. Thus, $J_{\chi}^{\langle O \rangle}$ is the contribution due to the $U(1)$-breaking condensation. The position for every jump is the same as Fig.~\ref{fig7}. Apart from a small range near $a_{\chi}=0$, for a given $a_{\chi}$, the magnitude of the current increases with the increase of $\alpha$, which means that the boundary system corresponding to large $\alpha$ is more sensitive to the response to Wilson line.

\section{Conclusions and Discussions}\label{Sec.5}
In this paper, we have studied the magnetic response of holographic superconductor in the Gauss-Bonnet gravity. Concretely we have studied the response to a Wilson line along the compact spatial direction of dual systems. As we know, the Gauss-Bonnet term effectively describes some kind of quantum correction in the bulk. According to AdS/CFT correspondence, such quantum correction maps to the $1/\mathcal{N}$ corrections in the boundary theory. It is interesting to investigate how much such quantum corrections change the whole picture. Our calculation shows that for a particular Gauss-Bonnet parameter, in the black brane background, the phase diagram and physical quantities , such as condensation and current, with different $(W,m)$ but equal $a_\chi-m/R$, are degenerate, while in the soliton phase such degeneracy is uplifted. Thus, we find that including the Gauss-Bonnet term does not modify the qualitative features observed in Einstein theory.

Although the Gauss-Bonnet term can not break the LP periodicity in black brane background, other physical quantities in the two holographic systems are undoubtedly modified or affected with the change of the Gauss-Bonnet parameter $\alpha$, which is equivalent to tuning the strength of quantum corrections. We have analyzed the impact of the Gauss-Bonnet parameter $\alpha$ on the response of Wilson line in detail. More specifically, different Gauss-Bonnet parameters affect the involved physical quantities with the following fashion.

The critical chemical potential $\mu_c$ defined by transition point for vanishing $a_\chi$ increases as one increases the Gauss-Bonnet parameter, which makes the condensation more difficult for vanishing magnetic flux. For a given $a_{\chi}R$, the smaller the Gauss-Bonnet parameter $\alpha$ is chosen, the bigger $\mu/\mu_c$ is required to trigger the phase transition. For the black brane case, the coherence length $\xi_0$ increases as we lower $\alpha$.

At fixed chemical potential compared to $\mu_c$ corresponding to each $\alpha$, the behaviors of condensation and current with respect to $a_\chi$ (or equivalently to Wilson line $W$) are much more different in the two systems with the change of the Gauss-Bonnet parameter $\alpha$. In the black brane case, we have found that the magnitudes of the condensation and current increase as one lowers the parameter $\alpha$, which means they are more sensitive to response to Wilson line for smaller $\alpha$. On the other hand, in the soliton background case, the magnitude of such two physical quantities increases with the increase of $\alpha$, indicating that the response to Wilson line is more insensitive for smaller $\alpha$.  Unlike the black brane case, there are jumps existing in the evolution of $\langle O\rangle$ and $J^{\langle O\rangle}_{\chi}$, which are due to the fact that the topological sectors labeled by $(W,m)$ enter the effective field theory of the boundary theory. And the position of the jumps moves to larger chemical potential with the increase of the Gauss-Bonnet parameter $\alpha$. Finally we point out that this work is done in the probe limit, the most direct improvement of the present analysis is to include the back reaction of the matter sector on the background geometry, but we expect the qualitative picture will not be changed.

\begin{acknowledgments}
We thank Song He and Shingo Takeuchi for helpful discussions,
LFL would like to thank A. Salvio for quick correspondence.
This work was supported in part by the National Natural Science
Foundation of China (No.10821504, No.10975168 and No.11035008), and
in part by the Ministry of Science and Technology of China under
Grant No. 2010CB833004. LFL was supported by the National Natural Science
Foundation of China with grant No.11205226 and China Postdoctoral
Science Foundation with grant No. 2012M510563.
\end{acknowledgments}


\end{CJK}
\end{document}